\documentclass{article}
\usepackage{amssymb,amsfonts,amsmath,mathtext,cite,enumerate,float}
\usepackage{setspace}
\usepackage{geometry} 
\begin{document}

\title{Calculation of thermal conductivity coefficients for magnetized neutron star}
\author{M. V. Glushikhina$^*$ and G. S. Bisnovatyi-Kogan$^{**}$}
\date{}
\maketitle
\thanks {Space Research Institute of Russian Academy of Sciences, Moscow,  Russia\\
$^*$E-mail: m.glushikhina@iki.rssi.ru\\
$^{**}$E-mail: gkogan@iki.rssi.ru\\}

\begin{abstract}
The coefficients that determine the electron heat transfer and diffusion in the crust of neutron stars are calculated on the basis of a solution of the Boltzmann equation with allowance for degeneracy.
\end{abstract}



\section{Introduction}\label{int}
Thermal  conductivity in the envelopes of neutron stars  plays crucial role in many aspects of evolution of these stars. Thermal conductivity is the basic quantity needed for calculating the  relationship between the internal temperature of a neutron star and its effective surface temperature, this relationship affects thermal evolution of the neutron star and its radiation spectra. We should know the transport properties of dense matter where electrons are strongly degenerate and form nearly ideal Fermi-gas. Under such conditions, electrons are usually most important  heat carriers. The present work is devoted to calculation of electron thermal conductivity coefficients under the given conditions on the basis of the  solution of Boltzmann equation. 
The magnetic field limits the motion of electrons in the direction perpendicular to the field lines and, since they are the main carriers of the heat transport, the thermal conductivity in this direction is highly suppressed, while remaining  unaffected along the field lines. In the  present work we obtain a  new approximation for coefficients of electron heat conductivity along and across magnetic field lines. 

\section{Boltzmann equations and transfer equations}
To calculate the transport coefficients   we use a Boltzmann equation with allowance for degeneracy and only two-body collisions.  We consider  an electron gas in the crystal lattice of  heavy nuclei and we take into account the interaction of the electrons with the nondegenerate nuclei and with one another.
The nuclear component of the matter in the crust is evidently in the crystal state and therefore the isotropic part of the distribution function $f_{N0}$ may differ from the Maxwellian distribution.  
If the mass of the $m_{N}$ nucleus is much greater, than the electron mass $m_{e}$ , the to the terms $\sim m_{e}/m_{N}$ the details of the distribution function $f_{N0}$ is unimportant, and the calculations can be made for arbitrary  $f_{N0}$. The transfer equations for the electron concentration, total momentum and energy in the two-component mixture of electrons and nuclei can be obtained in the usual manner~\cite{chap90}  from the Boltzmann equation for nonrelativistic electrons.~\cite{chap90, marsh90, bk64, bkl88}

\begin{equation}\label{electron number}
\frac{dn_{e}}{dt}+n_{e}\frac{\partial c_{0i}}{\partial r_i}+\dfrac{\partial }{\partial r_{i}}(n_{e}\langle   v_{i} \rangle)=0
\end{equation}

\begin{equation}\label{transport2}
\rho_{e}
\dfrac{dc_{0i}}{\mathit{dt}}=\dfrac 1 c\epsilon _{\mathit{ikl}}j_kB_l-\dfrac{\partial \Pi_{ik}}{\partial r_i}+\varepsilon(E_i+\dfrac 1 c\epsilon _{\mathit{ikl}}c_{\mathit{0k}}B_l)
\end{equation}

\begin{eqnarray}\label{energy}
\dfrac{3}{2}{kn_{e}}\dfrac{{dT}}{{dt}} - \dfrac{3}{2}kT\dfrac{\partial}{\partial r_{i}}(n_{e}\langle v_{i}\rangle)+\dfrac{\partial q_{ei}}{\partial r_{i}}+ \Pi_{ik}\dfrac{\partial c_{0i}}{\partial r_i}=j_{i}(E_{i}+\dfrac{1}{c}\epsilon_{ikl}c_{0k}B_{l}) - \rho_{e} \langle v_{i}\rangle\dfrac{dc_{0i}}{dt}
\end{eqnarray}

Where:
\begin{eqnarray}\label{pressure}
\begin{aligned}
p_{e} & =\dfrac{1}{3}n_{e}m_{e}\langle v^{2}\rangle  & \Pi_{ik}  & = nm\langle v_{i} v_{k}\rangle \\ 
q_{ i} & = \dfrac{1}{2}n m\langle v^{2}v_{i} \rangle  &  \langle v_{i} \rangle & = \dfrac{1}{n} \int f v_{i} dc_{i}\end{aligned}
\end{eqnarray} 
$c_{0i}$  is mass-average velocity, $v_{i}$ is the thermal velocity of the electrons.

\section{Derivation of general expressions for the transport coefficients}
The Boltzmann equation can be solved by the Chapmen-Enskog method of successive approximation\cite{chap90}. The zeroth approximation to the electron distribution function is found by equating to zero the collision integral:

\begin{equation}\label{dist_func}
\begin{aligned}
f_0 & =[1+\exp \frac{m_ev^2-2\mu}{2\mathit{kT}}]^{-1} & n_e & =D\int f_0\mathit{dv}_i
\end{aligned}
\end{equation}

Here, $\mu$ is chemical potential of electrons, k is Boltzmann's constant, T is the temperature, and $D=2m^3/(2\pi \hbar )^3$.

The nuclear distribution function in the zeroth approximation  $f_{N0}$ is assumed to be isotropic with respect to the  velocities and to depend on the local thermodynamic parameters; otherwise it can be arbitrary with the normalization:

\begin{equation}\label{nuclear_norm}\nonumber
n_N=\int f_{\mathit{N0}}\mathit{dc}_{Ni}
\end{equation}

Using ~(\ref{dist_func}) in~(\ref{electron number})-(\ref{pressure}), we obtain the zeroth approximation for the transfer equations. In this approximation $\langle v_{i} \rangle =0, q_{i} = 0, \Pi_{ik} = (\Pi_{e}+ \Pi_{N})\delta_{ik}$

\begin{equation}\label{el_num-den}
\begin{aligned}
n_e & =2(\frac{\mathit{kTm}_e}{2\pi ^2\hbar^2})^{3/2}G_{3/2}(x_0) & P_e & =2\mathit{kT}(\frac{\mathit{kTm}_e}{2\pi ^2\hbar ^2})^{3/2}G_{5/2}(x_0)
\end{aligned}
\end{equation} 

\begin{eqnarray}\label{fermi_integral}\nonumber
G_n(x_0)=\frac
1{\Gamma(n)}\int{ \dfrac{x^{n-1}dx}{1+exp(x-x_{0})} }\\ \nonumber
x_0=\frac{\mu }{\mathit{kT}}
\end{eqnarray}
In what follows, instead of $G_{n}(x_{0})$ we will write $G_{n}$ cause the argument is the same.
In the first approximation, we seek the function $f$ in the form:
\begin{equation}\label{chi}
f=f_0(1+\chi(1-f_0))
\end{equation}
We take deviation of the nuclear distribution function from zeroth approximation in the form:
\begin{equation}\label{chi_N}
f_N=f_{\mathit{N0}}(1+\chi_{N})
\end{equation}
$\chi$ is linear and admits representation of the solution in the form:
\begin{equation}\label{chi_a}
\begin{aligned}
\chi & = A_{i}\frac{\partial \ln T}{\partial r_{i}}-n_{e}D_{i}d_{i}\frac{G_{5/2}}{G_{3/2}} & \chi_N &=-A_{Ni}\frac{\partial \ln T}{\partial r_i} -n_{e}D_{Ni}d_{i}\frac{G_{5/2}}{G_{3/2}} 
\end{aligned}
\end{equation}

The functions $A_{i}, A_{Ni}$ and $D_{i}, D_{Ni}$ determine diffusion and heat transfer. Substituting ~(\ref{chi_a})  in the equation for $\chi$  we can obtain equations for $A_{i}, A_{Ni}, D_{i}, D_{Ni}$ \cite{dau90}

\begin{equation}\label{eq_sys}
\left\lbrace \begin{aligned} 
f_{0}(1-f_{0})(\dfrac{m_{e}v^{2}}{2kT} - \dfrac{5G_{5/2}}{2G_{3/2}})v_{i}  & = I_{ee}(A_{i})+I_{eN}(A_{i}) \\ 
\dfrac{1}{n_{e}} f_{0}(1-f_{0})v_{i}  & = I_{ee}(D_{i})+I_{eN}(D_{i}) \end{aligned} \right. \end{equation}

We seek solution of ~(\ref{eq_sys}) in the form of an expansion in polynomials $Q_{n}$  that are orthogonal with weight $f_{0}(1-f_{0})x^{3/2}$.
Where $Q_{n}$ - are analogous to Sonine polynomials~\cite{bkrom82}.
\begin{eqnarray}\label{sonine}
\begin{aligned}
Q_{0}(x) & = 1 &  x & = u^{2} = \frac{m_{e}}{2kT}v_{i}^{2} \\
Q_{1}(x)  &   = \frac{5G_{5/2}}{2G_{3/2}} - x
\end{aligned} 
\end{eqnarray}
We seek $A_{i}$ and $D_{i}$ in the form: 
\begin{eqnarray}\label{polin for A_D}
\begin{aligned}
A_{i}  & = (a_{0}Q_{0}+a_{1}Q_{1})v_{i} & D_{i}  & = (d_{0}Q_{0}+d_{1}Q_{1})v_{i}
\end{aligned}
\end{eqnarray}
 
 Multiplying ~(\ref{eq_sys}) by $DQ_{0}(x)u_{i}$ and $DQ_{1}(x)u_{i}$ and integrating with respect to $dc_{i}$, we obtain a system of equation for heat conductivity coefficients:
 \begin{eqnarray}\label{system}
\left\lbrace  \begin{aligned} 
0 & =3 i\dfrac{e B n_{e}}{m_{e} c} a_{0}+  a_{0}(a_{00}+b_{00}) +a_{1}(a_{01}+b_{01})\\
\dfrac{-15}{4} n_{e}(\dfrac{7G_{7/2}}{2G_{5/2}} - \dfrac{5G_{5/2}^{2}}{2G_{3/2}^{2}}) &  =i \dfrac{15 e B n_{e}}{4m_{e}c}a_{1} + a_{0}(a_{10}+b_{10}) + a_{1}(a_{11} + b_{11})
\end{aligned} \right.
\end{eqnarray}
 Here $a_{jk}$  $b_{jk}$ are matrix elements for collision integrals.

 \section{Finding of matrix elements}
 
 \begin{eqnarray}\label{aji} 
a_{jk}  = D^{2} \int  f_{0}f_{01}(1 - f_{0}^{'})(1- f_{01}^{'})Q_{j}(u^{2})u_{i}[Q_{k}(u^{2})  + Q_{k}(u_{1}^{2})u_{1i}- \nonumber  \\- Q_{k}(u^{'2})u_{i}^{'} - 
  Q_{k}(u_{1}^{'2}u_{1i}^{'})]g_{ee}W_{ee}(\theta, g_{ee})  d \Omega dc_{1i}dc_{i}
\end{eqnarray}
 
\begin{eqnarray}\label{b0j}
b_{0j} = D\int f_{0}f_{N0}(1-f_{0}^{'})Q_{j}(u^{2})u_{i}[u_{i} - u^{'}_{i} -  \dfrac{n_{e}}{n_{N}}(\dfrac{m_{e}}{m_{N}})^{1/2}(u_{Ni} - \nonumber \\ - u_{Ni}^{'})g_{eN}W_{eN}(\theta, g_{eN})d \Omega dc_{Ni} d c_{i}
\end{eqnarray} 
 
 \begin{eqnarray}\label{bjk}
 b_{jk} = D \int f_{0} f_{N0} (1-f_{0}^{'})Q_{j}(u^{2})u_{i}[Q_{k}(u^{2})u_{i} - \nonumber  \\  -Q_{k}(u^{'2})u_{i}^{'}]g_{eN}W_{eN}(\theta, g_{eN})d \Omega dc_{Ni} dc_{i}, \\ \nonumber  k\geq 1 
 \end{eqnarray}
 
 $W(\theta, g)d \Omega$ is the effective differential cross section for scattering of particles with relative velocity $g$ that is deflected through angle $\theta$ and after the collision lies in the solid angle $d\Omega$.
 
\section{Tensor of heat conductivity}
We can find $a_{0}$ and $a_{1}$ by solving ~(\ref{system}) and using following expressions:
\begin{equation}
\begin{aligned}
a_{0}  & = a_{0}^{1}+iBb_{0}^{1} &
a_{1}  & = a_{1}^{1}+iBb_{1}^{1} \\
c_{0}^{1} & = (a_{0}^{1})_{B=0} - a_{0}^{1}
 &  c_{1}^{1} & = (a_{1}^{1})_{B=0}-a_{1}^{1}
\end{aligned}
\end{equation}
 we can find coefficients of heat conductivity tensor ~\cite{bk64, bkl88, kl64}.

$\lambda_{ik} = \dfrac{5}{2}\dfrac{k^{2}Tn}{m}((a_{0}^{1} - a_{1}^{1})\delta_{ik} - \epsilon_{ikn}B_{n}(b_{0}^{1} - b_{1}^{1}) + B_{i}B_{k}(c_{0}^{1} - c_{1}^{1}))$

In previous works ~\cite{fl76, yau80} the following approximation was used:

\begin{equation}
\dfrac{\lambda\perp}{\lambda \parallel} = \dfrac{1}{1+(\omega \tau)^{2}}
\end{equation}
  
Considering two cases of strong and weak degeneracy, we obtain the following relations:

for weak degeneracy (nondegenerate):
\begin{equation}
\dfrac{\lambda\perp}{\lambda \parallel} = \dfrac{1}{1+\dfrac{17(\omega \tau)^{2}+2.76(\omega \tau)^{4}}{2.58 +0.39 (\omega \tau)^{2}}}
\end{equation}

for strong degeneracy:
\begin{equation}
\dfrac{\lambda\perp}{\lambda \parallel} = \dfrac{1}{1+\dfrac{0.31(\omega \tau)^{2}+0.55(\omega \tau)^{4}}{0.14 +0.6 (\omega \tau)^{2}}}
\end{equation}

here  in weak degeneracy  case $\tau = \dfrac{3 \sqrt{m_{e}}(kT)^{3/2}}{4 \sqrt{2 \pi} Z^{2} e^{4} n_{N} \Lambda}$ is relaxation time between electron-nucleus collisions  and in strong degeneracy case $ \tau = \dfrac{3 h^{3} n_{e}} {32  \pi^{2}  Z^{2} e^{4} n_{N} m_{e} \Lambda}$ and $\omega = \dfrac{eB}{m_{e}c}$ is electron cyclotron frequency. $\lambda_{\parallel} =  \dfrac{ k^{2}n_{e}^{2} T h^{3}}{32 m^{2}Z^{2}e^{4} n_{N} \Lambda}$ in the strong degeneracy case and $\lambda_{\parallel} = 4.4k \dfrac{n_{e}}{n_{N} \sqrt{2 \pi}}(\dfrac{kT}{m})^{1/2} \dfrac{(kT)^{2}}{Z^{2}e^{4}\Lambda}$ in the weak degeneracy case (nondegenerate).

In our work we obtain the more exact relation  for the heat conductivity coefficients along and across magnetic field lines by solving Boltzmann kinetic equation.

\section{Acknowledgements} \nonumber

The work of GSBK and MVG was partially supported
by the Russian Foundation for Basic Research grant 11-02-00602 and the Russian Federation President
Grant for Support of Leading Scientific Schools NSh-5440.2012.2.

The work of MVG was also partially supported by Russian
Federation President Grant for Support of Young Scientists MK-2918.2013.2.

\end{document}